\documentclass{PoS}
\usepackage{bm}
\usepackage{bbm}
\usepackage{IEEEtrantools}
\usepackage{amsmath}
\usepackage{pstricks}
\DeclareMathOperator{\tr}{tr}
\usepackage{ctable}
\allowdisplaybreaks[1]

\title{Low-energy limit of the $\bm{O(4)}$ quark-meson model}

\ShortTitle{Low-energy limit of the O(4) quark-meson model}

\author{\speaker{J\"urgen Eser} and Florian Divotgey\\
        Institut f\"ur Theoretische Physik, Johann Wolfgang Goethe-Universit\"at,\\ 
        Max-von-Laue-Stra\ss e 1, D-60438 Frankfurt am Main, Germany\\
        E-mail: \email{eser@th.physik.uni-frankfurt.de},
        \email{fdivotgey@th.physik.uni-frankfurt.de}}
        
\author{Mario Mitter\\
        Department of Physics, Brookhaven National Laboratory,\\ 
        Upton, NY 11973, USA\\
        E-mail: \email{mitter@bnl.gov}}

\abstract{We study the generation of low-energy couplings induced 
by quantum fluctuations within the $O(4)$-symmetric quark-meson model.
To this end, we compute the functional renormalization group flow of the
linearly realized quark-meson model including higher-derivative 
interactions and subsequently transform the resulting effective 
action into a nonlinear effective pion action. The latter is referred 
to as the low-energy limit of the $O(4)$ quark-meson model.
The present study may be considered as a preparatory work for the 
dynamical generation of low-energy couplings from functional QCD 
fluctuations in order to determine meaningful renormalization scales 
for purely pionic models.}

\FullConference{The 9th International workshop on Chiral Dynamics\\
		17-21 September 2018\\
		Durham, NC, USA}

\begin{document}

\section{Introduction}

The predominant feature of the strong interaction at low energies 
is the chiral $SU(N_{f})\times SU(N_{f})$ symmetry, where $N_{f}$ 
represents the number of dynamical quark flavors. From the spontaneous
breakdown of this symmetry arise $N_{f}^{2} - 1$ (pseudo) Nambu-Goldstone 
bosons (pNGBs), which, in the case of $N_{f} = 2$, are identified with 
the three pion fields. The chiral symmetry and its breaking is therefore 
the main guiding principle for the construction of effective field theories 
(EFTs) describing the hadronic low-energy regime of the strong interaction. 

The most prominent EFT is given by chiral perturbation theory (ChPT)
\cite{Gasser:1983yg, Gasser:1984gg}, which is based on a nonlinear 
realization of chiral symmetry. It relies on a simultaneous expansion 
of the generating functional of QCD in terms of pion momenta and
quark masses. The associated Lagrangian of ChPT involves (momentum-dependent)
pion self-interactions parametrized by low-energy constants.

So-called linear sigma models (LSMs) constitute another interesting family 
of EFTs for the strong interaction. In contrast to ChPT, they are 
constructed from a linear realization of chiral symmetry, treating 
the pNGB fields and their chiral partners on equal footing. These 
LSMs also have a longstanding tradition and are 
frequently used, e.g., for nonzero-temperature and nonzero-chemical 
potential applications. The $O(4)$ LSM, emerging from 
the relation $SU(2)\times SU(2) \sim O(4)$, is one of the 
simplest examples in this family, which considers the three pion fields 
$\vec{\pi}$ and their chiral partner, the $\sigma$ meson. Taking also 
constituent quarks into account, it is extended to the $O(4)$ quark-meson 
model (QMM), the primary subject of investigations in the following.

The presented work is motivated by the question whether quantum
fluctuations of the (had\-ronic) degrees of freedom within the $O(4)$ 
QMM and related EFTs, such as extended LSMs \cite{Parganlija:2010fz, 
Parganlija:2012fy}, are able to reproduce the low-energy couplings of 
QCD? To address this question, we compute fluctuation-induced higher-derivative 
interactions in the linearly realized QMM and transform the infrared
effective action into a nonlinearly realized effective pion action by
restricting the dynamics to the respective vacuum manifold.

In a first (exploratory) approach \cite{Eser:2018jqo}, we studied 
the scaling behavior of higher-derivative pion self-interactions within the 
$O(4)$ QMM using the functional renormalization group (FRG) formalism. This
investigation was initialized by the tree-level computation of the 
low-energy couplings in an extended LSM \cite{Divotgey:2016pst}. 
The subsequent study in Ref.\ \cite{Divotgey:2019xea} elaborated on this 
first analysis by including momentum-dependent $\sigma\pi$ as well as 
$\sigma$ self-interactions into the FRG flow. This second work also allows 
for a first conclusion regarding appropriate renormalization scales for 
purely pionic EFTs as obtained from the FRG. Here, we briefly review 
latest results of these studies.

\section{Methods}

In Secs.\ \ref{sec:QMM} and \ref{sec:FRG}, we introduce the $O(4)$ QMM
and the related truncation within the FRG formalism including  
higher-derivative interactions. The next section \ref{sec:pionaction} 
shows the corresponding low-energy limit, i.e., the nonlinearly realized 
effective pion action derived from the linear QMM. This action contains
(momentum-dependent) pion self-interactions, which are parametrized 
by the low-energy couplings of the QMM.

\subsection{Quark-meson model: Lagrangian}
\label{sec:QMM}

The $O(4)$ QMM is constructed from the four-dimensional vector
\begin{equation}
	\varphi = \begin{pmatrix} \vec{\pi} \\ \sigma 
	\end{pmatrix}, \label{eq:phi}
\end{equation}
which contains the three pion fields $\vec{\pi}$ and the isoscalar $\sigma$ 
field. The Lagrangian of the QMM (on Minkowski spacetime) reads
\begin{equation}
	\mathcal{L}_{\text{QMM}} = \frac{1}{2}\left(\partial_{\mu}
	\varphi\right)\cdot\partial^{\mu}\varphi - \frac{m_{0}^{2}}{2}
	\varphi\cdot\varphi - \frac{\lambda}{4}\left(\varphi\cdot\varphi\right)^{2} 
	+ h_{\text{ESB}}\sigma + \bar{\psi}\left(i\gamma^{\mu}\partial_{\mu} 
	- y\Phi_{5}\right)\psi, \label{eq:lagrangianQMM}
\end{equation}
with
\begin{equation}
	\Phi_{5} = \sigma t_{0} + i\gamma_{5}\vec{\pi}\cdot\vec{t}.
	\label{eq:phi5}
\end{equation}
In Eq.\ (\ref{eq:lagrangianQMM}), we introduced the Yukawa coupling 
$y$ of the mesonic fields to the quarks ($\bar{\psi}$, $\psi$) as well
as the parameter $h_{\text{ESB}}$, which leads to an explicit breaking
of the $O(4)$ invariance of the model (besides the spontaneous breaking).
The generators in Eq.\ (\ref{eq:phi5}) are defined as $t_{0} = \mathbbmss{1}/2$ 
and $\vec{t} = \vec{\tau}/2$, where $\vec{\tau}$ are the three Pauli matrices.

\subsection{Quark-meson model: Truncation with higher-derivative interactions}
\label{sec:FRG}

The FRG implements the integration over quantum fluctuations in terms
of momentum shells. It works with the scale-dependent effective average 
action $\Gamma_{k}$, where the subscript $k$ denotes the flowing infrared 
(IR) cutoff in momentum space. The action $\Gamma_{k}$ interpolates 
between the renormalized classical action at a specific ultraviolet (UV) 
cutoff $\Lambda$, $k\rightarrow \Lambda$, and the quantum effective action 
$\Gamma$ in the IR, $k\rightarrow 0$.

The scaling behavior of the effective average action $\Gamma_{k}$ 
is determined by the Wetterich flow equation \cite{Wetterich:1992yh}
\vspace{8pt}
\begin{equation}
	\partial_{k}\Gamma_{k} = \frac{1}{2}\tr\left[
	\partial_{k} R_{k}\left(
	\Gamma^{(2)}_{k} + R_{k}\right)^{-1}\right]. 
	\label{eq:Wetterich} \vspace{5pt}
\end{equation}
The function $R_{k}$ in the above equation denotes the IR regulator 
and $\Gamma_{k}^{(2)}$ is the scale-dependent two-point function, i.e., 
the second functional derivative of $\Gamma_{k}$ with respect to the 
(classical) fields.

We solve Eq.\ (\ref{eq:Wetterich}) by choosing a truncation for the 
Euclidean QMM, cf.\ the Lagrangian (\ref{eq:lagrangianQMM}) given in 
the previous section, which involves complete sets of higher-derivative 
couplings of order $\mathcal{O}(\varphi^{4},\partial^{2})$ and 
$\mathcal{O}(\varphi^{4},\partial^{4})$,
\begin{IEEEeqnarray}{rCl}
	\Gamma_{k} & = & \int_{x}\bigg\lbrace \frac{Z_{k}}{2}
	\left(\partial_{\mu}\varphi\right) \cdot 
	\partial_{\mu}\varphi + U_{k}(\rho) - h_{\mathrm{ESB}}\sigma 
	+ \bar{\psi}\left(Z_{k}^{\psi}\gamma_{\mu}
	\partial_{\mu} + y_{k} \Phi_{5}\right)\psi \nonumber\\
	& & \qquad +\, C_{2,k}\left(\varphi \cdot 
	\partial_{\mu}\varphi\right)^{2}
	+ Z_{2,k}\, \varphi^{2}\left(\partial_{\mu}\varphi\right)
	\cdot \partial_{\mu}\varphi \nonumber\\
	& & \qquad -\, C_{3,k} \left[\left(\partial_{\mu}\varphi\right)
	\cdot \partial_{\mu}\varphi\right]^{2}
	- C_{4,k} \left[\left(\partial_{\mu}\varphi\right)
	\cdot \partial_{\nu}\varphi\right]^{2}
	- C_{5,k} \, \varphi \cdot \left(\partial_{\mu}\partial_{\mu}\varphi\right)
	\left(\partial_{\nu}\varphi\right) \cdot \partial_{\nu}\varphi \nonumber\\
	& & \qquad -\, C_{6,k} \, \varphi^{2} \left(\partial_{\mu}\partial_{\nu}\varphi\right)
	\cdot \partial_{\mu}\partial_{\nu}\varphi
	- C_{7,k} \left(\varphi \cdot \partial_{\mu}\partial_{\mu}\varphi\right)^{2}
	- C_{8,k} \, \varphi^{2}\left(\partial_{\mu}\partial_{\mu}\varphi\right)^{2}
	\bigg\rbrace .\label{eq:truncation}
\end{IEEEeqnarray}
This truncation was developed in our first work \cite{Eser:2018jqo} on 
this topic. Apart from the derivative couplings $C_{2,k}$ and $Z_{2,k}$ as well as 
$C_{i,k}$, $i = 3,\ldots , 8$, this ansatz also includes an $O(4)$-symmetric 
scale-dependent effective potential $U_{k}(\rho)$, where $\rho = \varphi 
\cdot \varphi = \vec{\pi}^{2} + \sigma^{2}$, and wave-function 
renormalization factors for the bosonic and fermionic fields, $Z_{k}$ and 
$Z_{k}^{\psi}$, respectively. Furthermore, a scale dependence of the Yukawa
coupling $y_{k}$ is taken into account.

The flow equations for the scale-dependent constituents of truncation
(\ref{eq:truncation}) are obtained from Eq.\ (\ref{eq:Wetterich}) and 
its functional derivatives, see Refs.\ \cite{Eser:2018jqo} and 
\cite{Divotgey:2019xea} for details.

\subsection{Low-energy limit of the quark-meson model: Effective pion action}
\label{sec:pionaction}

Restricting the dynamics of the original four-dimensional Euclidean field
space (spanned by the fields $\vec{\pi}$ and $\sigma$) to the 
vacuum manifold associated with the spontaneous symmetry breaking, 
i.e., the three-sphere $S^{3}$, yields the low-energy limit of 
the $O(4)$ QMM (the quark fields are dropped as they do not influence 
vacuum-to-vacuum amplitudes computed from the effective action at tree 
level; the fermionic quantum fluctuations are already integrated into 
the higher-derivative couplings).

The transition to the nonlinear low-energy limit is formally 
denoted as
\begin{equation}
	\Gamma_{k}\left[\tilde{\sigma}, \tilde{\vec{\pi}}, 
	\tilde{\bar{\psi}}, \tilde{\psi}\right] \rightarrow 
	\Gamma_{k}\left[\tilde{\Pi}\right], \label{eq:transition}
\end{equation}
where $\tilde{\cdot}$ are the respective renormalized fields,
\begin{equation}
	\tilde{\sigma} = \sqrt{Z_{k}^{\pi}}\sigma, \qquad
	\tilde{\vec{\pi}} = \sqrt{Z_{k}^{\pi}}\vec{\pi}, \qquad
	\tilde{\bar{\psi}} = \sqrt{Z_{k}^{\psi}}\bar{\psi}, \qquad
	\tilde{\psi} = \sqrt{Z_{k}^{\psi}}\psi,
\end{equation}
with $Z_{k}^{\pi}$ as the effective pion wave-function renormalization 
in the two-point function,
\begin{equation}
	Z_{k}^{\pi} = Z_{k} + 2 \sigma^{2} Z_{2,k} - 2\sigma^{2}
	p^{2} \left(C_{6,k} + C_{8,k}\right).
\end{equation}
This factor $Z_{k}^{\pi}$ obviously differs from $Z_{k}$ as soon as
the $\sigma$ field (and the external momentum $p$) acquires a nonzero 
value. The corresponding renormalized higher-derivative couplings 
in truncation (\ref{eq:truncation}) read
\begin{equation}
	\tilde{C}_{i,k} = \frac{C_{i,k}}{(Z_{k}^{\pi})^{2}}, \quad
	i = 1, \ldots , 8, \qquad \tilde{Z}_{2,k} = \frac{Z_{2,k}}
	{(Z_{k}^{\pi})^{2}},
\end{equation}
where $C_{1,k}$ is identified with the momentum-independent quartic
interaction in the effective potential $U_{k}$. In Eq.\ (\ref{eq:transition}), 
the fields $\tilde{\Pi}^{a}$, $a = 1,2,3$, represent the properly renormalized 
nonlinear pNGBs and we kept the symbol $\Gamma_{k}$ for the resulting effective 
pion action.

After choosing stereographic coordinates on the three-sphere, the
effective pion action for the $O(4)$ QMM turns out to be (see again Ref.\ 
\cite{Divotgey:2019xea} for technical details)
\begin{IEEEeqnarray}{rCl}
	\Gamma_{k} & = & \int_{x} \biggl\lbrace \frac{1}{2}
	\left(\partial_{\mu}\tilde{\Pi}_{a}\right)\partial_{\mu}\tilde{\Pi}^{a} 
	+ \frac{1}{2}\tilde{\mathcal{M}}^{2}_{\Pi,k}\, \tilde{\Pi}_{a}\tilde{\Pi}^{a}
	- \tilde{\mathcal{C}}_{1,k}\left(\tilde{\Pi}_{a}\tilde{\Pi}^{a}
	\right)^{2} + \tilde{\mathcal{Z}}_{2,k}\, \tilde{\Pi}_{a}\tilde{\Pi}^{a}
	\left(\partial_{\mu}\tilde{\Pi}_{b}\right)\partial_{\mu}\tilde{\Pi}^{b} \nonumber\\
	& & \quad\quad -\, \tilde{\mathcal{C}}_{3,k}\left[\left(\partial_{\mu}\tilde{\Pi}_{a}\right)
	\partial_{\mu}\tilde{\Pi}^{a}\right]^{2}
	- \tilde{\mathcal{C}}_{4,k}\left[\left(\partial_{\mu}\tilde{\Pi}_{a}
	\right)\partial_{\nu}\tilde{\Pi}^{a}\right]^{2}
	- \tilde{\mathcal{C}}_{5,k}\, \tilde{\Pi}_{a}\left(\partial_{\mu}\partial_{\mu}
	\tilde{\Pi}^{a}\right)\left(\partial_{\nu}\tilde{\Pi}_{b}\right)\partial_{\nu}
	\tilde{\Pi}^{b} \nonumber\\
	& & \quad\quad -\, \tilde{\mathcal{C}}_{6,k}\, \tilde{\Pi}_{a}\tilde{\Pi}^{a}
	\left(\partial_{\mu}\partial_{\nu}\tilde{\Pi}_{b}\right)\partial_{\mu}\partial_{\nu}\tilde{\Pi}^{b}
	- \tilde{\mathcal{C}}_{8,k}\, \tilde{\Pi}_{a}\tilde{\Pi}^{a}
	\left(\partial_{\mu}\partial_{\mu}\tilde{\Pi}_{b}\right)\partial_{\nu}\partial_{\nu}\tilde{\Pi}^{b}
	\biggr\rbrace .\label{eq:finalaction}
\end{IEEEeqnarray}
The low-energy couplings in the nonlinear model (\ref{eq:finalaction}) 
are functions of the corresponding couplings in the linear QMM,
\begin{IEEEeqnarray}{rCl}
	\tilde{\mathcal{C}}_{1,k} & = & \frac{\tilde{\mathcal{M}}^{2}_{\Pi,k}}{8f^{2}_{\pi}}, \qquad
	\tilde{\mathcal{Z}}_{2,k} = -\frac{1}{4f^{2}_{\pi}}, \nonumber\\
	\tilde{\mathcal{C}}_{3,k} & = & \tilde{C}_{3,k} - \tilde{C}_{5,k} + \tilde{C}_{7,k}
	+ 2\bigl(\tilde{C}_{6,k} + \tilde{C}_{8,k}\bigr), \qquad
	\tilde{\mathcal{C}}_{4,k} = \tilde{C}_{4,k}, \nonumber \\
	\tilde{\mathcal{C}}_{5,k} & = & 2\bigl(\tilde{C}_{6,k}+\tilde{C}_{8,k}\bigr), \qquad
	\tilde{\mathcal{C}}_{6,k} = -\, \tilde{C}_{6,k}-\tilde{C}_{8,k}, \qquad
	\tilde{\mathcal{C}}_{8,k} = \frac{1}{2}\bigl(\tilde{C}_{6,k} 
	+ \tilde{C}_{8,k}\bigr). \label{eq:LECrelations}
\end{IEEEeqnarray}
To be more precise, the couplings $\tilde{\mathcal{C}}_{1,k}$ and 
$\tilde{\mathcal{Z}}_{2,k}$ are geometrical constants, which means that
they are exclusive functions of the pion decay constant $f_{\pi}$ (the 
radius of the three-sphere) and the (squared) pion mass $\tilde{
\mathcal{M}}^{2}_{\Pi,k}$,
\begin{equation}
	\tilde{\mathcal{M}}^{2}_{\Pi,k} = \frac{\tilde{h}_{\text{ESB}}}{f_{\pi}}
	= \frac{h_{\text{ESB}}}{\sqrt{Z_{k}^{\pi}}}\frac{1}{f_{\pi}}.
\end{equation}
The couplings $\tilde{\mathcal{C}}_{5,k}$, $\tilde{\mathcal{C}}_{6,k}$,
and $\tilde{\mathcal{C}}_{8,k}$ in the nonlinear model appear
to be linearly dependent. The couplings $\tilde{\mathcal{C}}_{2,k}$ and
$\tilde{\mathcal{C}}_{7,k}$ vanish.

\section{Numerical results}
\label{sec:results}

The numerical values of the low-energy couplings of the effective
pion action (\ref{eq:finalaction}) are computed by solving the 
FRG flow of the $O(4)$ QMM according to truncation (\ref{eq:truncation})
and subsequently using the relations (\ref{eq:LECrelations}) to transform 
the result into the nonlinear model.
\begin{figure}[b!]
	\centering
		\includegraphics{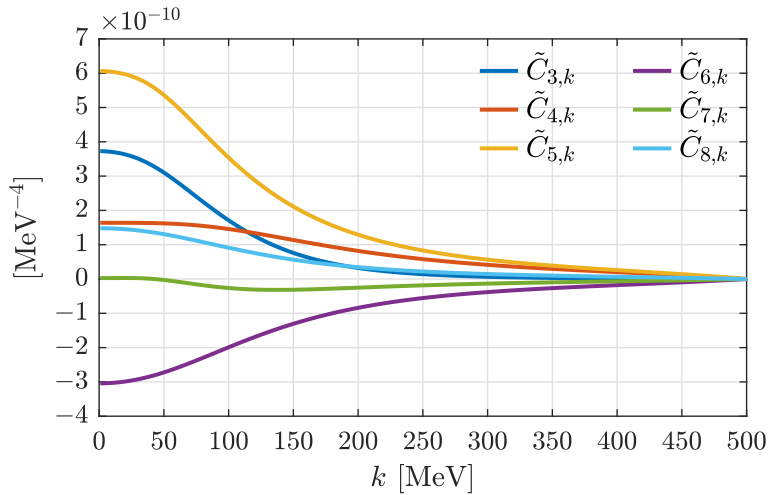}
	\caption{Higher-derivative couplings. Scale evolution 
	of the renormalized couplings $\tilde{C}_{i,k}$, 
	$i = 3, \ldots , 8$, of the $O(4)$ QMM (\ref{eq:truncation}),
	cf.\ Ref.\ \cite{Divotgey:2019xea}; $k_{\mathrm{IR}} = 1\ \mathrm{MeV}$.}
	\label{fig:cp4}
\end{figure}

The flow of the QMM is initialized at a hadronic cutoff scale
of $\Lambda = 500\ \text{MeV}$. At this scale, the higher-derivative
couplings $C_{2,k}$ and $Z_{2,k}$ as well as $C_{i,k}$, $i = 3,\ldots ,
8$, are set to zero [Ref.\ \cite{Eser:2018jqo} confirmed that this
is a reasonable assumption for the derivative couplings of order
$\mathcal{O}(\partial^{4})$]. Hence, these interactions are 
solely dynamically generated during the integration process. 
All other concrete starting values, e.g., for the effective 
potential and the Yukawa interaction, can be found 
in Ref.\ \cite{Divotgey:2019xea}. These values are tuned 
such that, in the IR, the pions, the $\sigma$ meson, and the
quarks have a mass of 138.5 MeV, 509.4 MeV, and 296.8 MeV, respectively.
The pion decay constant $f_{\pi}$ amounts to 93.8 MeV in the IR limit.
\begin{figure}[t!]
	\centering
		\includegraphics{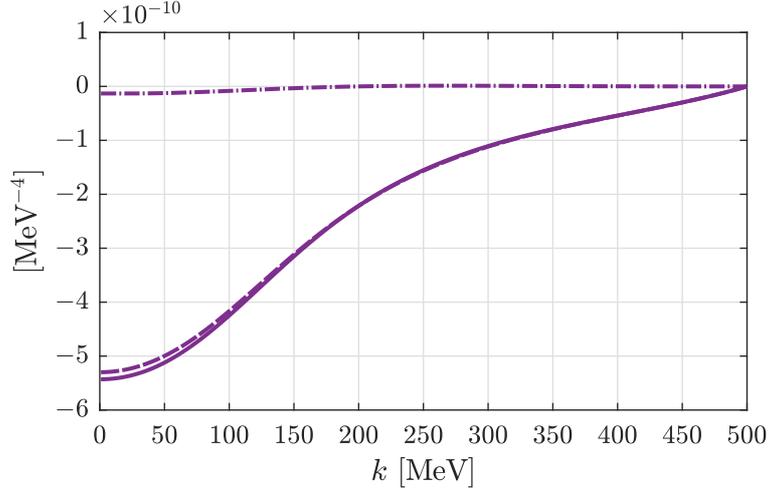}
	\caption{Low-energy coupling $\tilde{\mathcal{C}}_{3,k}$. Scale 
	evolution of the renormalized low-energy coupling $\tilde{\mathcal{C}}_{3,k}$,
	which is qualitatively representative for the derivative couplings
	in the effective pion action (\ref{eq:finalaction}), cf.\ Ref.\ 
	\cite{Divotgey:2019xea}. The total evolution (solid line) is
	decomposed into bosonic (dash-dotted line) and fermionic (dashed line) 
	loop contributions; $k_{\mathrm{IR}} = 1\ \mathrm{MeV}$.}
	\label{fig:cnonlinear}
\end{figure}

Figure \ref{fig:cp4} shows the generation of the derivative 
couplings of order $\mathcal{O}(\partial^{4})$ during the 
flow of the linear QMM [we do not show the derivative couplings 
$\tilde{C}_{2,k}$ and $\tilde{Z}_{2,k}$, which are irrelevant 
for the low-energy couplings, cf.\ relations (\ref{eq:LECrelations})].  
The IR values of the higher-derivative couplings are listed in the first 
part of Table \ref{tab:LECs} (``Linear model''). They differ from the ones 
quoted in our first calculation \cite{Eser:2018jqo}
due to the influence of the momentum-dependent 
$\sigma\pi$ and $\sigma$ self-interactions on the FRG flow.

\ctable[caption={Higher-derivative couplings and
the related low-energy (derivative) couplings.
Table from Ref.\ \cite{Divotgey:2019xea}.},botcap,
label=tab:LECs,pos=b!]{l c l c}{}{
\FL
	\multicolumn{2}{c}{\textbf{Linear model}} & \multicolumn{2}{c}{\textbf{Nonlinear model}}
\ML
	$\tilde{C}_{2}\ [1/f_{\pi}^{2}]\times 10$ & $-0.88$ & \multicolumn{2}{c}{$\ldots$} \\
	$\tilde{Z}_{2}\ [1/f_{\pi}^{2}]\times 10$ & $-2.30$ & 
	$\tilde{\mathcal{Z}}_{2}\ [1/f_{\pi}^{2}]\times 10$ & $-2.50$ \\
	$\tilde{C}_{3}\ [1/f_{\pi}^{4}]\times 10^{2}$ & $2.88$  & 
	$\tilde{\mathcal{C}}_{3}\ [1/f_{\pi}^{4}]\times 10^{2}$ & $-4.20$ \\
	$\tilde{C}_{4}\ [1/f_{\pi}^{4}]\times 10^{2}$ & $1.27$  & 
	$\tilde{\mathcal{C}}_{4}\ [1/f_{\pi}^{4}]\times 10^{2}$ & $1.27$ \\
	$\tilde{C}_{5}\ [1/f_{\pi}^{4}]\times 10^{2}$ & $4.69$  & 
	$\tilde{\mathcal{C}}_{5}\ [1/f_{\pi}^{4}]\times 10^{2}$ & $-2.41$ \\
	$\tilde{C}_{6}\ [1/f_{\pi}^{4}]\times 10^{2}$ & $-2.35$ & 
	$\tilde{\mathcal{C}}_{6}\ [1/f_{\pi}^{4}]\times 10^{2}$ & $1.21$ \\
	$\tilde{C}_{7}\ [1/f_{\pi}^{4}]\times 10^{2}$ & $0.02$  & \multicolumn{2}{c}{$\ldots$} \\
	$\tilde{C}_{8}\ [1/f_{\pi}^{4}]\times 10^{2}$ & $1.14$  & 
	$\tilde{\mathcal{C}}_{8}\ [1/f_{\pi}^{4}]\times 10^{2}$ & $-0.60$
\LL
}

It is clear that the transition to the nonlinear effective
pion action is only meaningful at energy-momentum scales at which 
the non-pNGB degrees of freedom have decoupled from the flow,
i.e., they have become too massive. Nevertheless, we can use 
Eq.\ (\ref{eq:LECrelations}) to determine the low-energy couplings 
of the nonlinear model (\ref{eq:finalaction}) at all scales.

Thus, as an example for the low-energy couplings $\tilde{\mathcal{C}}_{i,k}$,
$i \in \lbrace 3,4,5,6,8 \rbrace$, all of which exhibit the same 
qualitative behavior \cite{Divotgey:2019xea}, Fig.\ 
\ref{fig:cnonlinear} presents the scale evolution of the coupling 
$\tilde{\mathcal{C}}_{3,k}$. Additionally, the evolution 
of this coupling is decomposed into bosonic and fermionic loop 
contributions. Apparently, the latter dominate the FRG flow and
the bosonic loop contributions are only important at
scales roughly below 150 MeV.

Finally, the IR-limit values of the low-energy couplings 
of the action (\ref{eq:finalaction}) are also collected
in Table \ref{tab:LECs}; see the second part (named ``Nonlinear model'').
The vanishing couplings in the nonlinear case are indicated by three dots.

\section{Summary and outlook}

We have computed the low-energy limit of the $O(4)$ QMM 
within the FRG framework. This low-energy limit is given
by a nonlinearly realized effective pion action, whose 
interaction terms are parametrized by low-energy couplings.
These couplings were determined from the flow of the corresponding
higher-derivative interactions in the linear QMM.

We found that the FRG flow is drastically dominated by the
fermionic fluctuations \cite{Divotgey:2019xea}, even for relatively
small scales below 100 MeV, cf.\ Fig.\ \ref{fig:cnonlinear}. 
Moreover, the pNGB fluctuations are only relevant at scales roughly 
below 150 MeV; a result, which is consistent with recent functional 
QCD publications, see, e.g., Refs.\ \cite{Mitter:2014wpa, Braun:2014ata, 
Cyrol:2017ewj, Paris-Lopez:2018vjc, Alkofer:2018guy}. These 
findings suggest renormalization scales of around 50 -- 100 MeV for 
the effective pion action as obtained from the FRG.

The present studies are understood as a preparatory work for
the derivation of low-energy couplings from functional QCD, i.e.,
from quark-gluon fluctuations initialized at scales of order
$10\ \text{GeV}$. Such an analysis will follow in the near
future.

\acknowledgments{The authors are grateful to D.H.\ Rischke for
his support. J.E.\ thanks the German National Academic 
Foundation and HIC for FAIR for funding. M.M.\ was supported 
by the DFG grant MI 2240/1-1 and the U.S.\ Department of Energy 
under contract de-sc0012704.}

\bibliographystyle{JHEP}
\bibliography{references_proceedings}

\end{document}